\documentclass[twocolumn,prl,showpacs]{revtex4}

\usepackage{graphicx}
\usepackage{dcolumn}
\usepackage{bm}
\usepackage{amsmath}

\begin{document}

\preprint{Helical edge}
\title{Anomalous Andreev bound state in non-centrosymmetric superconductors }
\author{ Yukio Tanaka$^{1}$, Yoshihiro Mizuno$^{1}$,  
Takehito Yokoyama$^{2}$,  Keiji Yada$^{1}$, and Masatoshi Sato$^{3}$}
\affiliation{$^1$Department of Applied Physics,
Nagoya University, Nagoya, 464-8603,
Japan \\
$^2$ Department of Physics, Tokyo Institute of Technology, Tokyo, 152-8551,
Japan \\
$^3$ Institute for Solid State Physics, University of Tokyo, Chiba 277-8581, Japan
}
\date{\today}

\begin{abstract}
We study edge states of non-centrosymmetric superconductors where
 spin-singlet $d$-wave pairing mixes with spin-triplet  $p$ (or
 $f)$-wave one by spin-orbit coupling. 
For $d_{xy}$-wave pairing, the obtained Andreev bound state has an
anomalous dispersion as compared to conventional helical edge modes. 
A unique topologically protected time-reversal invariant Majorana 
bound state  appears at the edge. 
The charge conductance in the non-centrosymmetric superconductor junctions
 reflects the anomalous structures of the dispersions, particularly the
 time-reversal invariant Majorana bound state is manifested as a
 zero bias conductance peak.
\end{abstract}

\maketitle



%

%



Recently, physics of non-centrosymmetric (NCS) 
superconductors is one of the important issues in condensed 
matter physics \cite{Bauer,Zheng,Frigeri,Interface}.  
One of the remarkable features in NCS superconductors
is that due to the 
broken inversion symmetry, superconducting pair potential becomes a mixture of 
spin-singlet even-parity and spin-triplet odd-parity \cite{Gorkov}. 
Due to the mixture of spin-singlet and spin-triplet pairings, 
several novel properties 
such as the large upper critical field are expected \cite{Frigeri,Fujimoto1}. \par
Up to now, there have been several studies about 
superconducting profiles of NCS superconductors 
\cite{Frigeri,Fujimoto1,Yanase,Linder,Iniotakis,Eschrig,Tanaka,Sato2009,Yip}. 
In these works, pairing symmetry of NCS superconductors has been 
mainly assumed to be $s+p$-wave.
However, in a strongly correlated system, this assumption is not 
valid anymore. 
Microscopic calculations have shown that $d_{x^{2}-y^{2}}$-wave spin-singlet pairing mixes with $f$-wave pairing based on the Hubbard model near half filling \cite{NCS-theory}. 
Also, a possible pairing symmetry of superconductivity generated at heterointerface LaAlO$_3$/SrTiO$_3$ \cite{Interface} has been studied 
based on  a similar model \cite{Yada}. 
It has been found that the gap function consists 
of spin-singlet $d_{xy}$-wave component and spin-triplet 
$p$-wave one \cite{Yada}.  
Therefore, now, it is  a challenging issue to reveal novel properties specific to  $d_{xy}+p$ or $d_{x^{2}-y^{2}}+f$-wave pairing. \par
The generation of Andreev bound state(ABS) at the surface 
or interface is a significantly important feature  specific to unconventional 
pairing since ABS directly manifests itself in the tunneling spectroscopy. 
Actually, for $d_{xy}$-wave pairing, zero energy dispersionless ABS appears \cite{ABS}. 
The presence of ABS has been verified by tunneling experiments of high-T$_{c}$ 
cuprate \cite{TK95} as a zero bias conductance peak (ZBCP). 
For NCS superconductors, 
when $p$-wave pair potential is larger than $s$-wave one, 
it has been shown that 
ABS is generated at its edge as helical edge modes similar to those in quantum spin Hall system\cite{Iniotakis,Eschrig,Tanaka,Sato2009}. 
Several new features of  spin transport
stemming from these helical edge modes 
have been also predicted \cite{Eschrig,Tanaka,Sato2009,Yip}.
However, there has been no theory about  ABS 
in $d_{xy}+p$- or $d_{x^{2}-y^{2}}+f$-wave pairing in NCS superconductors. 
Since tunneling spectroscopy via ABS \cite{ABS} is a powerful method to identify pairing symmetry and mechanism 
of unconventional superconductors \cite{TK95}, it is quite important and 
interesting 
to clarify ABS and resulting tunneling conductance for 
$d_{xy}+p$-wave and $d_{x^{2}-y^{2}} + f$-wave pairings.  \par
%
In this Letter, we investigate 
ABS and  tunneling conductance $\sigma_{C}$ in normal metal / NCS superconductor junctions.  
For both $d_{xy}+p$-wave and $d_{x^{2}-y^{2}} + f$-wave 
cases, new types of ABS are obtained.
In particular, for $d_{xy}+p$-wave case, 
due to the Fermi surface splitting by 
spin-orbit coupling, 
a single branch of topologically stable Majorana bound state appears.
Recently, to search for Majorana fermions
is one of the hottest issues in condensed matter
physics\cite{Wilczek,Majorana1}.  
In stark contrast to the other Majorana fermions, the
present one preserves time-reversal symmetry.
From this difference,
the ``time-reversal invariant (TRI) Majorana bound state'' has a
peculiar flat dispersion. 
It shows a
unique ZBCP in $\sigma_{C}$ depending on the spin-orbit coupling. 
Therefore, the experimental identification is feasible. \par

We start with the Hamiltonian of NCS superconductor
\begin{eqnarray}
\check{H}_{S} = \left( {\begin{array}{*{20}c}
{\hat H\left( {\bf k} \right)} & {\hat \Delta \left( {\bf k} \right)} \\
{ - \hat \Delta ^ * \left( { - {\bf k}} \right)} & { - \hat H^ * \left( 
{ - {\bf k}} \right)} \\
\end{array}} \right)
\label{Hamiltonian}
\end{eqnarray}
with
$\hat{H}({\bm k})
=\xi_{{\bm k}} + \bm{V}(\bm{k}) \cdot \hat{\bm{\sigma}}
$,
$\bm{V}({\bm k})=\lambda (\hat{\bm{x}}k_{y}-\hat{\bm{y}}k_{x})
$,
$\xi_{\bm{k}}=\hbar^2 {\bm k}^{2}/(2m) - \mu$.
Here, $\mu$, $m$,
$\hat{\bm{\sigma}}$ and $\lambda$ denote
chemical potential, effective mass,
Pauli matrices and coupling constant of Rashba spin-orbit
interaction, respectively \cite{Frigeri}.
The pair potential $\hat{\Delta}(\bm{k})$
is given by 
$
\hat{\Delta}(\bm{k}) = 
[\bm{d}(\bm{k})\cdot \hat{{\bm \sigma}}]i\hat{\sigma}_{y}
+i \psi(\bm{k})\hat{\sigma}_{y}. 
$
Due to the spin-orbit coupling, the spin-triplet component $\bm{d}(\bm{k})$
is aligned with the polarization vector of the Rashba spin orbit
coupling, $\bm{d}(\bm{k}) || \bm{V}(\bm{k})$\cite{Frigeri}. 
Then, the triplet component is $\bm{d}(\bm{k})=\Delta_{t}f({\bm k})
(\hat{\bm{x}}k_{y}-\hat{\bm{y}}k_{x})/ k$ 
with $k=\sqrt{{\bm k}^2}$ 
while singlet component reads
$\psi(\bm{k})=\Delta_{s}f({\bm k})$ 
with $\Delta_{t} \geq 0$ and 
$\Delta_{s} \geq 0$. 
$f(\bm{k})$ is given by 
$f(\bm{k})=2k_{x}k_{y}/k^2$ for 
$d_{xy}+p$-wave 
and $f(\bm{k})=(k_{x}^{2} -k_{y}^{2})/k^2$ 
for $d_{x^{2}-y^{2}} + f$-wave.\cite{dxy}
The superconducting gaps are 
$\Delta_{1}=|\bar{\Delta}_{1}(\bm{k})|$ 
and 
$\Delta_{2}=|\bar{\Delta}_{2}(\bm{k})|$ 
for the 
two spin-split band 
with 
$\bar{\Delta}_{1}({\bm k})=(\Delta_{t}+\Delta_{s})f({\bm k})$ 
and 
$\bar{\Delta}_{2}({\bm k})=(\Delta_{t}-\Delta_{s})f({\bm k})$, 
respectively, in homogeneous state \cite{Iniotakis}. 

Let us  consider a wave function including ABS localized at the
surface. 
Consider a two-dimensional semi-infinite superconductor
on $x>0$ where the surface is located at $x=0$. 
The corresponding wave function 
%
is given by \cite{Tanaka}
\begin{eqnarray}
\Psi_{S}(x)=
[c_{1}^{+}\psi_{1}^{+}\exp(iq^{+}_{1x}x) 
+ c_{1}^{-}\psi_{1}^{-}\exp(-iq^{-}_{1x}x)
\nonumber\\
+ c_{2}^{+}\psi_{2}^{+}\exp(iq^{+}_{2x}x) 
+ c_{2}^{-}\psi_{2}^{-}\exp(-iq^{-}_{2x}x) ]\exp(ik_{y}y), 
\label{wavefunction}
\\
q^{\pm}_{1(2)x}
=k^{\pm}_{1(2)x} \pm \frac{ k_{1(2)}}{k^{\pm}_{1(2)x}}
\sqrt{\frac{E^{2}-[ \bar \Delta_{1(2)}(\bm{k}_{1(2)}^{\pm})]^{2}}
{\lambda^{2} + 2\hbar^{2}\mu /m }},
\nonumber
\end{eqnarray}
with 
$k^{+}_{1(2)x}=k^{-}_{1(2)x}=\sqrt{k_{1(2)}^2-k_y^2}$ for $|k_{y}|
\leq k_{1(2)}$
and $k^{+}_{1(2)x}=-k^{-}_{1(2)x}=i\sqrt{k_{y}^{2}-k_{1(2)}^2}$ 
for $|k_{y}|>k_{1(2)}$, and 
$\bm{k}_{1(2)}^{\pm}=(\pm k_{1(2)x}^{\pm},k_{y})$. 
Here, $k_{1}$ and $k_{2}$ are the Fermi wavenumbers for the smaller and larger
Fermi surface given by $-m\lambda/\hbar^{2} + \sqrt{
(m\lambda/\hbar^{2})^{2} + 2m\mu/\hbar^{2} }$ and $m
\lambda/\hbar^{2} + \sqrt{ (m\lambda /\hbar^{2})^{2} +
2m\mu/\hbar^{2} }$, respectively.
The wave functions are
given by $^T\psi_{1}^{\pm} =\left(
1,-i\alpha_{1\pm}^{-1},i\alpha_{1\pm}^{-1}\Gamma_{1\pm},\Gamma_{1\pm} \right)$ 
and $^T\psi_{2}^{\pm} =\left(1,i\alpha_{2\pm}^{-1},
i \alpha_{2\pm}^{-1}\Gamma_{2\pm},-\Gamma_{2\pm} \right)$
with
\begin{eqnarray}
\Gamma_{1(2)\pm}=
\frac{\bar{\Delta}_{1(2)}(\bm{k}_{1(2)}^{\pm})}
{ E \pm \sqrt{E^2 - [\bar{\Delta}_{1(2)}(\bm{k}_{1(2)}^{\pm}) ] ^2 } },
\end{eqnarray}
and $\alpha_{1(2)\pm}=(\pm k^{\pm}_{1(2)x}-ik_{y})/k_{1(2)}$.
$E$ is the
quasiparticle energy measured from the Fermi energy.
\par
Postulating $\Psi_{S}(x)=0$ at $x=0$, we can determine the
ABS. We consider the case for $|k_{y}| < k_{2}$. 
We first focus on the ABS for $d_{xy} + p$-wave case. 
For $\Delta_{t}>\Delta_{s}$,  
the dispersion $\varepsilon_{b}$ of ABS is given by 
\begin{eqnarray}
\displaystyle
\varepsilon_{b} =
\left\{
\begin{array}{ll}
\frac{\pm 2 \Delta_{t}\gamma  \sqrt{(k_{1}^{2}-k_{y}^{2})(k_{2}^{2}-k_{y}^{2})(k_{y}^{2}-\eta^{2}k_{1}^{2})}}
{(k_{1}+k_{2})k_{1}k_{2}}
& k_{c}<|k_{y}| \leq k_{1} \\
0 & k_{1} < |k_{y}|
\end{array}
\right. 
\label{bound}
\end{eqnarray}
with $\gamma=(k_{1}/k_{2} + k_{2}/k_{1}) + 
(\Delta_{s}/\Delta_{t})(k_{2}/k_{1} - k_{1}/k_{2})$, 
$\eta=[\Delta_{t}(1-k_{1}/k_{2}) 
+ \Delta_{s}(1 + k_{1}/k_{2})]/ 
\{\Delta_{t}[1+(k_{1}/k_{2})^{2}] 
+ \Delta_{s}[1 -(k_{1}/k_{2})^{2}]\}$, 
$k_{c}=k_{1}\sqrt{\Delta_{t}(1-k_{1}/k_{2})+\Delta_{s}}/
\sqrt{\Delta_{t} + \Delta_{s}[1-(k_{1}/k_{2})^{2}]}$.
On the other hand, for $\Delta_{s}>\Delta_{t}$, the resulting
$\varepsilon_{b}$ is given by
$\varepsilon_{b} =0. $
The dispersion $\varepsilon_{b}$ of ABS changes drastically at
$\Delta_{s}=\Delta_{t}$, 
where one of the energy gaps, i.e. $\Delta_{2}$, becomes zero. 
It should be remarked that the present ABSs do not break the 
time reversal symmetry. \par
The resulting $\varepsilon_{b}$ is plotted for various cases in 
Fig. 1 with $\Delta_{0}=\Delta_{s}+\Delta_{t}$. 
For convenience, we introduce dimensionless constant  
$\beta= 2m \lambda/(\hbar^{2} k_{f})$ with $k_{f}=\sqrt{2m\mu /\hbar^{2}}$. 
We also plot $\Delta_{1}$ and $\Delta_{2}$. 
Both $\Delta_{1}$ and $\Delta_{2}$ become zero at $k_{y}=0$. 
At $|k_{y}|=k_{2}$, $\Delta_{2}$ is always zero. 
However,  $\Delta_{1}$ then becomes zero only for $\beta=0$. 
First, we look at the $\Delta_{t}>\Delta_{s}$ case. 
For $\Delta_{s}=0$ with $\beta=0$, 
$\varepsilon_{b}=\pm c k_{y}$ with some constant $c$ for small $k_{y}$ 
(curve $a$ in Fig. 1(A)) as shown in the case of $s + p$-wave
pairing\cite{Linder,Iniotakis,Eschrig,Tanaka,Sato2009,Yip} 
 since $\eta=0$ is satisfied. This type of ABS is called helical edge mode 
\cite{Tanaka,Sato2009,Qi}. 
However, this  condition is satisfied 
only for $\Delta_{s}=0$ and $\beta=0$. 
In fact, $\varepsilon_{b}$ near $k_{y}=0$ becomes absent 
in general as shown in curves $a$ in Figs. 1(B), (D) and (E). 
At $k=\pm k_{c}$, $\varepsilon_{b}$ coincides with $\pm \Delta_{2}$. 
For nonzero $\beta$, 
$\varepsilon_{b}$ becomes exactly zero for $|k_{y}|>k_{1}$ 
as shown in curves $a$ in Figs. 1(D) and (E). 
The present line shapes of $\varepsilon_{b}$ are completely different from 
those of $s+p$-wave superconductors. 
On the other hand, for $\Delta_{s}>\Delta_{t}$, 
$\varepsilon_{b}=0$ for any $k_{y}$ similar to the case of 
spin-singlet $d_{xy}$ or spin-triplet $p_{x}$-wave pairing \cite{ABS,TK95}. 

We notice here that the zero energy bound state for $|k_{y}|>k_{1}$ 
is a Majorana bound state.
The wave function for the zero energy edge state 
$\Psi_{m}(k_{y})$ can be written as  
 $^{T}\Psi_{m}(k_{y}) =
 (u_{1}(k_{y}),u_{2}(k_{y}),v_{1}(k_{y}),v_{2}(k_{y}))$ where
\begin{eqnarray}
u_{1}(k_{y})=-i\sigma v_{2}(k_{y})
=\frac{(\alpha f_{1}-\beta_{1}f_{2})\exp(ik_{y}y-i\frac{\pi}{4})}
{\sqrt{\sigma \alpha}} 
\\ 
u_{2}(k_{y})=i\sigma v_{1}(k_{y})
=\frac{(f_{1}+\beta_{2}f_{2})\exp(ik_{y}y-i\frac{\pi}{4})}
{\sqrt{\sigma \alpha}}
\end{eqnarray}
with $\alpha=(k_{y}-\sqrt{k_{y}^{2}-k_{1}^{2}})/k_{1}$, 
$\beta_{1}=(\alpha k_{y}/k_{2} +1)$, 
$\beta_{2}=(\alpha + k_{y}/k_{2})$ and $\sigma={\rm sgn}(k_{y})$.  
The functions $f_{1}$ and $f_{2}$ decays exponentially as a function of $x$ and are even function of $k_{y}$. 
The Bogoliubov quasiparticle creation operator for this state is constructed in the usual way as $\gamma^\dagger(k_{y}) 
= u_1(k_y) c_\uparrow^\dagger(k_y) + u_2(k_y) c_\downarrow^\dagger(k_y) 
+ v_1(k_y) c_\uparrow(-k_y) + v_2(k_y) c_\downarrow(-k_y)$. 
Since $u_{1}(k_{y})=v_{1}^{*}(-k_{y})$ and 
$u_{2}(k_{y})=v_{2}^{*}(-k_{y})$ are satisfied, it is possible to verify that 
$\gamma^{\dagger}(k_{y})=\gamma(-k_{y})$.  
This means the generation of Majorana bound state at the edge for 
$|k_{y}| > k_{1}$. For $\Delta_{s}>\Delta_{t}$, 
a similar Majorana bound state also 
appears for  $|k_{y}| > k_{1}$. 
On the other hand, for $|k_{y}| \leq k_{1}$, Majorana bound state has double 
branches and it is reduced to be conventional zero energy ABS. 
\par
Unlike Majorana fermions studied before \cite{Wilczek, Majorana1}, 
the present single Majorana bound 
state is realized with time reversal symmetry. 
The TRI Majorana bound state has the following three characteristics.
a) It has a unique flat dispersion: To be consistent with the
time-reversal invariance, the single branch of zero mode should be
symmetric under $k_y\rightarrow -k_y$.
Therefore, by taking into account the particle-hole symmetry as well,
the flat dispersion is required. 
On the other hand, the conventional time-reversal breaking Majorana bound state has a linear dispersion.
b) The spin-orbit coupling is
necessary to obtain the TRI Majorana bound state.
Without spin-orbit coupling, the TRI Majorana bound state vanishes. 
c) The TRI Majorana bound state is topologically stable under small
deformations of the Hamiltonian (\ref{Hamiltonian}).

\begin{figure}[htb]
\begin{center}
\scalebox{0.8}{
\includegraphics[width=9.5cm,clip]{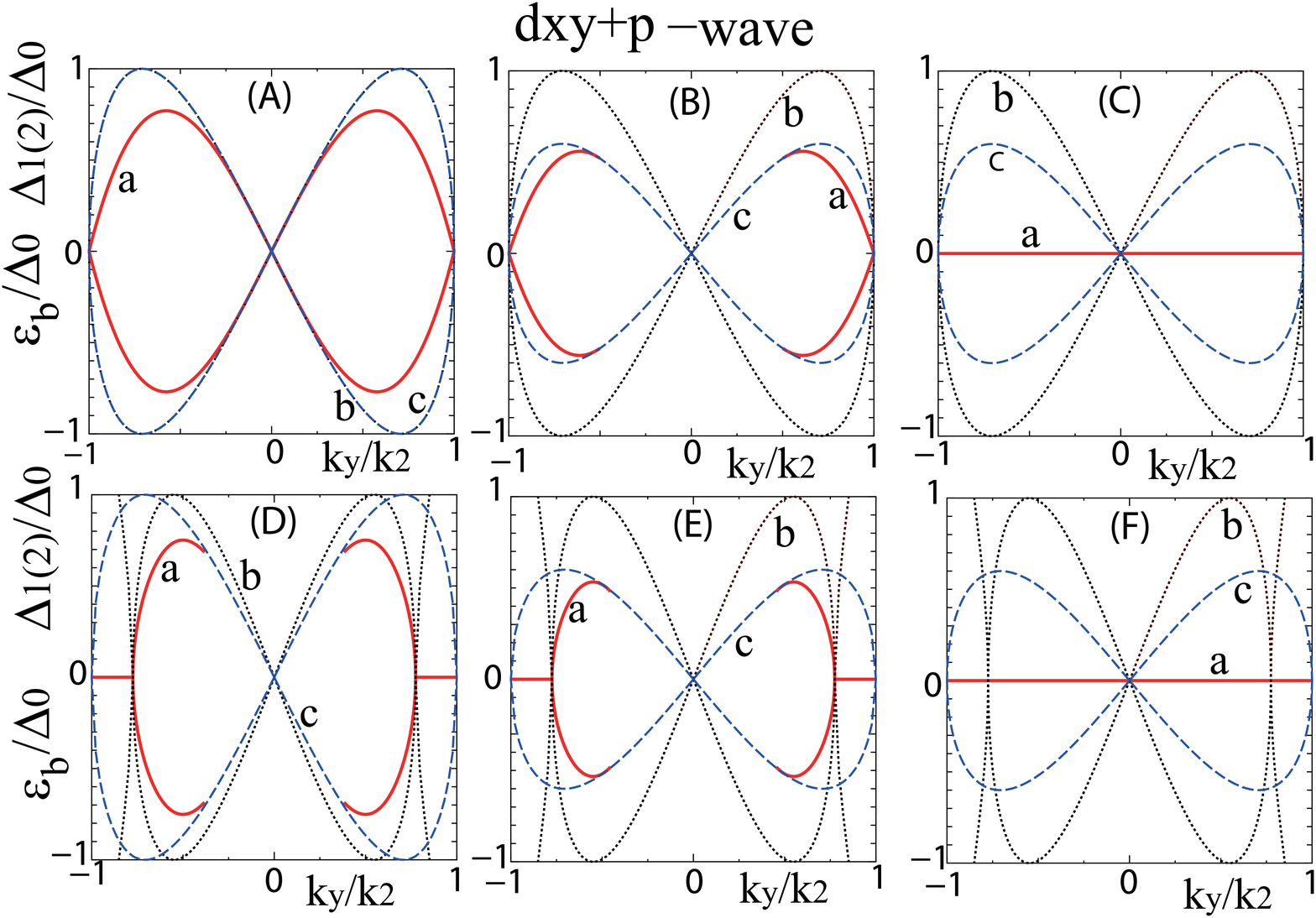}
}
\end{center}
\caption{(Color online) Andreev bound state 
$\varepsilon_{b}$, effective pair potentials 
for each Fermi surface $\Delta_{1}$ and $\Delta_{2}$ 
are plotted 
for $d_{xy} + p$-wave case as a function of 
$k_{y}/k_{2}$. 
$\beta=0$ for panels A, B and  C.  $\beta=0.5$ for panels D, E and F. 
$\Delta_{t}=\Delta_{0}$, $\Delta_{s}=0$ for A and D.  
$\Delta_{t}=0.8\Delta_{0}$, $\Delta_{s}=0.2\Delta_{0}$ for B and E. 
$\Delta_{t}=0.2\Delta_{0}$, $\Delta_{s}=0.8\Delta_{0}$, for C and F. 
In all panels, curves a (solid line),  b (dotted line) and 
c (dashed line) denote $\varepsilon_{b}/\Delta_{0}$, 
$\Delta_{1}/\Delta_{0}$ and 
$\Delta_{2}/\Delta_{0}$, respectively.  }
\label{fig:1}
\end{figure}
We also calculate ABS for $d_{x^{2}-y^{2}} + f$-wave case. 
In this case, ABS exists only for $\Delta_{s}<\Delta_{t}$.  
In Fig. 2, $\varepsilon_{b}$ is plotted similarly to Fig. 1. 
As a reference, corresponding $\varepsilon_{b}$ is also shown for $s+p$-wave 
case. Helical edge modes around $k_{y}=0$ exist and 
$\varepsilon_{b}$ is absorbed into continuum levels for $|k_{y}|>k_{1}$. 
These features are similar to those of $s+p$-wave case. 
However, the number of crossing points of $\varepsilon_{b}$ is five for  $d_{x^{2}-y^{2}} + f$-wave case 
reflecting the complex ${\bm k}$-dependence of the pair potential. 
The overall line shapes of $\varepsilon_{b}$ (curve $a$) in Fig. 2(A) is 
significantly different from corresponding $\varepsilon_{b}$ (curve $a$) in Fig. 2(B). \par
%
\begin{figure}[htb]
\begin{center}
\scalebox{0.8}{
\includegraphics[width=7.5cm,clip]{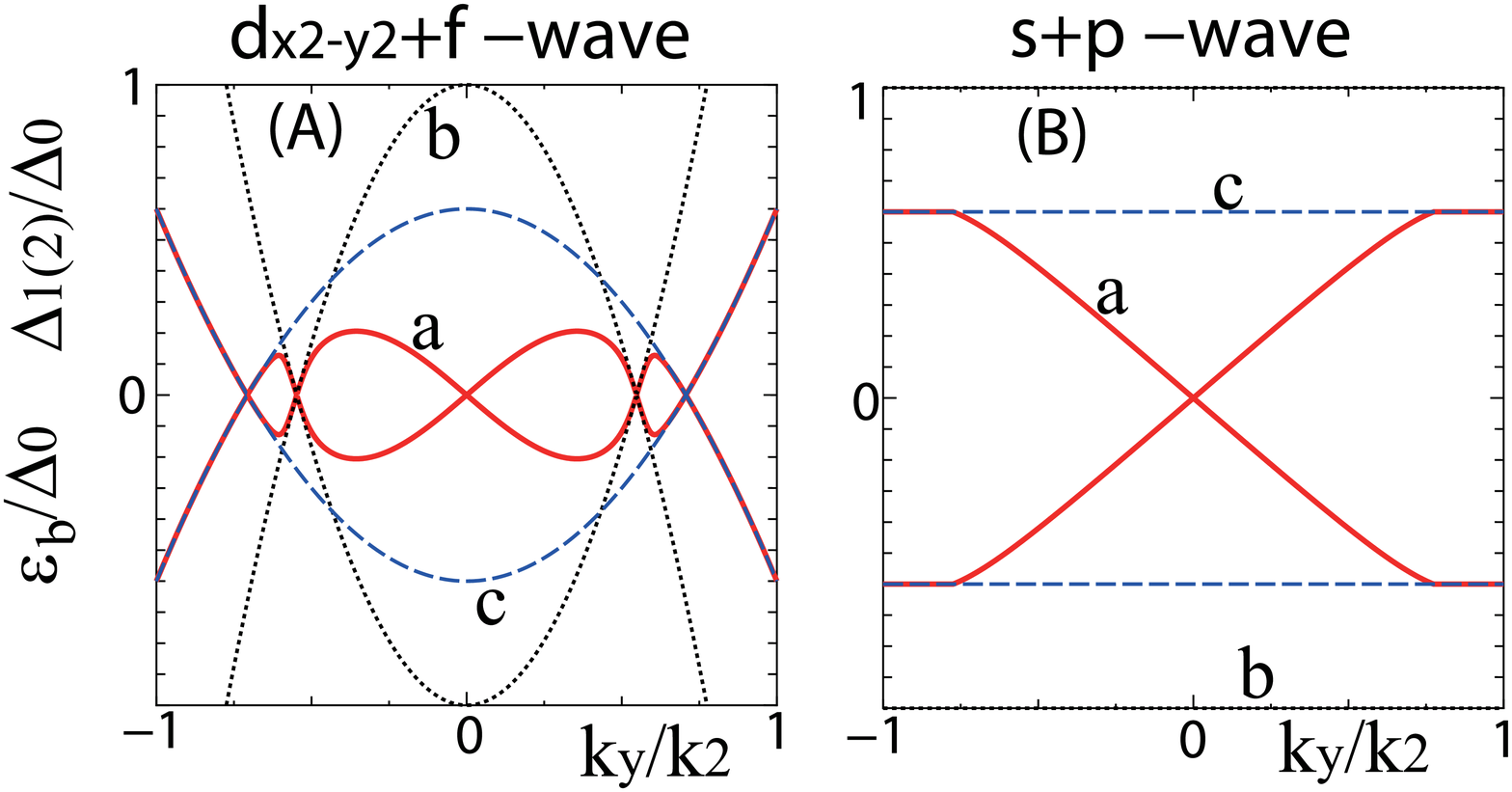}
}
\end{center}
\caption{(Color online) 
Similar plots to Fig. 1 with $\beta=0.5$ 
for $d_{x^{2}-y^{2}} + f$-wave (A) 
and $s + p$-wave case (B) 
with 
$\Delta_{t}=0.8\Delta_{0}$, $\Delta_{s}=0.2\Delta_{0}$. 
In all panels, curves a(solid line), b(dotted line) and c(dashed line) 
denote $\varepsilon_{b}/\Delta_{0}$, $\Delta_{1}/\Delta_{0}$ and 
$\Delta_{2}/\Delta_{0}$, respectively. }
\label{fig:2}
\end{figure}
It is very interesting to clarify how the above novel types of 
ABS are reflected in the charge transport property  
\cite{Iniotakis}.
The Hamiltonian $\check{H}_{N}$ in a normal metal is given by putting 
$\hat{\Delta}(\bm{k})=0$ and $\lambda=0$ in $\check{H}_{S}$.
We assume an insulating barrier at $x=0$
expressed by a delta-function potential $U \delta(x)$. 
The wave function for spin $\gamma=(\uparrow,\downarrow)$ in
the normal metal $\Psi_{N}(x)$ 
is given by
\begin{eqnarray}
\Psi_{N}(x)\!\!&=&\!\!\exp(ik_{Fy}y)
[(\psi_{i\gamma}+\sum_{\rho=\uparrow,\downarrow}a_{\gamma,\rho}\psi_{a\rho})\exp(ik_{Fx}x)
\nonumber\\
&&+ \sum_{\rho=\uparrow,\downarrow}
b_{\gamma,\rho}\psi_{b\rho}\exp(-ik_{Fx}x)]
\end{eqnarray}
with
$^T\psi_{i\uparrow}$$=$
$^T\psi_{b\uparrow}$
$=$$\left(1,0,0,0 \right)$, 
$^T\psi_{i\downarrow}$=$^T\psi_{b\downarrow}$
$=$$\left(0,1,0,0 \right)$, 
$^T\psi_{a\uparrow}$
$=$$\left(0,0,1,0 \right)$, and 
$^T\psi_{a\downarrow}$
$=$$\left(0,0,0,1 \right)$. 
The corresponding $\Psi_{S}(x)$ is given by Eq. (\ref{wavefunction}). 
The coefficients $a_{\gamma,\rho}$ and $b_{\gamma,\rho}$ are 
determined by the boundary condition 
$\Psi_{N}(0)=\Psi_{S}(0)$, and 
$\hbar \check{v}_{Sx}\Psi_{S}(0)-\hbar \check{v}_{Nx}\Psi_{N}(0)
=-2iU\check{\tau}_{3}\Psi_{S}(0)$ 
with $\hbar \check{v}_{S(N)x}=\partial \check{H}_{S(N)}/\partial k_{x}$, 
and diagonal matrix $\check{\tau}_{3}$ given by 
$\check{\tau}_{3}={\rm diag}(1,1,-1,-1)$. 

The quantity of interest is the angle averaged charge conductance 
$\sigma_{C}$ given by 
\begin{eqnarray}
\sigma_{C}=
\frac{\int^{\pi/2}_{-\pi/2} f_{C}(\phi) d\phi }
{\int^{\pi/2}_{-\pi/2} f_{NC}(\phi) d\phi}, 
\label{average}
\\
f_{C}(\phi)
=[2 + \sum_{\gamma,\rho} (\mid a_{\gamma,\rho} \mid^{2} -
\mid b_{\gamma,\rho} \mid^{2} ) ]
\frac{\cos \phi}{2}, 
\end{eqnarray}
where $f_{NC}(\phi)$ denotes the angle resolved 
charge conductance in the normal state with 
$\hat{\Delta}({\bm k})=0$. 
Here, $\phi$ denotes the injection angle measured from the 
normal to the interface with 
$\sin \phi =k_{y}/k_{f}$. To characterize transparency of 
the junction interface, 
we introduce dimensionless constant $Z=2mU/\hbar^{2}k_{f}$. \par

We plot bias voltage $eV=E$ dependence of $\sigma_{C}$ for 
$d_{xy}+p$-wave case in Fig. 3 for various $Z$.  
First we concentrate on low transparent junction with $Z=5$ by changing the 
value of $\Delta_{s}$ and $\Delta_{t}$. 
At $\Delta_{t}=\Delta_{s}$, one of the energy gap of the Fermi surface 
closes corresponding to the quantum phase transition. 
Then, the resulting $\sigma_{C}$ has a gradual change  from the quantum critical point.
\begin{figure}[htb]
\begin{center}
\scalebox{0.8}{
\includegraphics[width=9.0cm,clip]{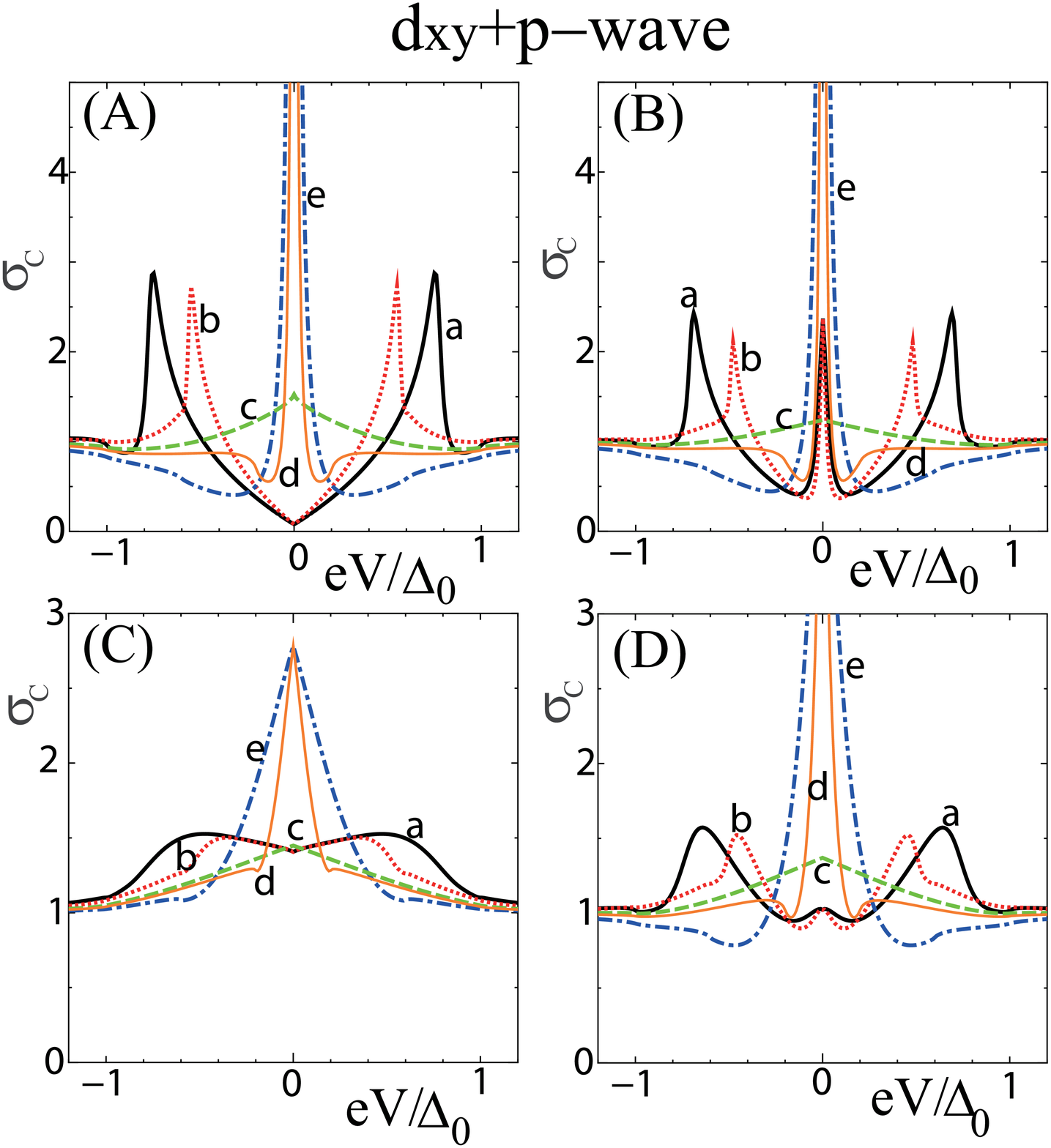}
}
\end{center}
\caption{(Color online) 
Tunneling conductance $\sigma_{C}$ for 
$d_{xy}+p$-wave case. 
A: $Z=5$, $\beta=0$, B: $Z=5$, $\beta=0.5$,  
C: $Z=1$, $\beta=0.5$, D: $Z=2$, $\beta=0.5$.  
a(solid line): $\Delta_{t}=\Delta_{0}$, $\Delta_{s}=0$, 
b(dotted line): $\Delta_{t}=0.8\Delta_{0}$, $\Delta_{s}=0.2\Delta_{0}$, 
c(dashed line): $\Delta_{t}=0.5\Delta_{0}$, $\Delta_{s}=0.5\Delta_{0}$, 
d(thin solid line): $\Delta_{t}=0.4\Delta_{0}$, $\Delta_{s}=0.6\Delta_{0}$, 
and 
e(dot-dashed line): $\Delta_{t}=0.2\Delta_{0}$, $\Delta_{s}=0.8\Delta_{0}$.  }
\label{fig:3}
\end{figure}
For the case without spin-orbit coupling ($\beta=0$) with 
$\Delta_{t}>\Delta_{s}$, $\sigma_{C}$ has a gap like 
structure around zero bias due to the absence of Majorana bound state as shown 
in Figs. 1(A) and 1(B). 
For $\Delta_{s}>\Delta_{t}$, ZBCP appears 
reflecting the zero energy ABS\cite{TK95}.  
In the presence of spin-orbit coupling, 
$\sigma_{C}$ always has a ZBCP 
independent of the ratio of $\Delta_{s}$ and $\Delta_{t}$ as shown in
Fig. 3(B). 
For $\Delta_{t}>\Delta_{s}$, the ZBCP originates from purely TRI 
Majoana bound state.  
The width of the ZBCP for $\Delta_{t}>\Delta_{s}$ is enhanced with the 
increase of $\beta$, since the region of $k_{y}$ where 
the TRI Majorana bound state exists is expanded with $\beta$. 
For $\Delta_{s}>\Delta_{t}$, both the 
conventional ABS and TRI Majorana bound state contribute to the 
formation of ZBCP.  
\begin{figure}[htb]
\begin{center}
\scalebox{0.8}{
\includegraphics[width=7.7cm,clip]{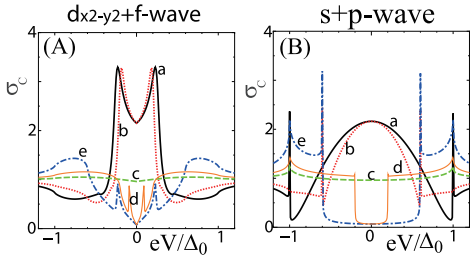}
}
\end{center}
\caption{(Color online) 
Tunneling conductance $\sigma_{C}$ with $\beta=0.5$ and 
$Z=5$. 
A: $d_{x^{2}-y^{2}} + f$-wave and 
B: $s+p$-wave. 
a(solid line): $\Delta_{t}=\Delta_{0}$, $\Delta_{s}=0$, 
b(dotted line): $\Delta_{t}=0.8\Delta_{0}$, $\Delta_{s}=0.2\Delta_{0}$, 
c(dashed line): $\Delta_{t}=0.5\Delta_{0}$, $\Delta_{s}=0.5\Delta_{0}$, 
d(thin solid line): $\Delta_{t}=0.4\Delta_{0}$, $\Delta_{s}=0.6\Delta_{0}$, 
and 
e(dot-dashed line): $\Delta_{t}=0.2\Delta_{0}$, $\Delta_{s}=0.8\Delta_{0}$.   }
\label{fig:4}
\end{figure}
We also plot corresponding $\sigma_{C}$ for high ($Z=1$) 
and intermediate $(Z=2)$ transparent junctions. 
For $\Delta_{t}>\Delta_{s}$, $\sigma_{C}$ has a broad dip-like structure around $eV=0$ for $Z=1$, while it is slightly enhanced around $eV=0$ for $Z=2$
(curves $a$ and $b$ in Figs. 3(C) and (D)).  
On the other hand, for $\Delta_{s}>\Delta_{t}$, $\sigma_{C}$ always 
has a ZBCP (curves $d$ and $e$ in Figs.3(C) and (D)). 
The presence of TRI Majorana bound state gives a clear ZBCP 
with the increase of $Z$. 
As a reference, the tunneling conductance $\sigma_{C}$ for $d_{x^{2}-y^{2}}+f$-wave and $s+p$-wave cases are plotted in Fig. 4 for $Z=5$. 
ABS exists only for $\Delta_{s}<\Delta_{t}$. 
The $\sigma_{C}$ for $d_{x^{2}-y^{2}}+f$-wave has a ZBCP splitting 
reflecting the complex dispersion $\varepsilon_{b}$ shown in Fig. 4(A). 
On the other hand, for $s+p$-wave case, $\sigma_{C}$ has a broad ZBCP 
shown in Fig. 4(B). Summarizing Figs. 3 and 4, $\sigma_{C}$ for each paring state are qualitatively different from each other, which can be used to identify these pairings.  
\par
In conclusion, we have studied the ABS and 
resulting charge transport 
for $d_{xy} +p$-wave and $d_{x^{2}-y^{2}} +f$-wave superconductors. 
We find that the obtained dispersion of ABS in both cases have  
an anomalous structure. 
For $d_{xy} +p$-wave case, a
novel TRI Majorana  bound state is generated due to the
spin-orbit coupling.
The resulting charge conductance can 
serve as a guide to identify the TRI Majorana bound state
and paring symmetry of 
NCS superconductors by tunneling spectroscopy. 
\par
This work is partly supported by the Sumitomo Foundation (M.S.) and the
Grant-in-Aids for Scientific Research No.\ 22103005 (Y.T. and M.S.), 
No.\ 20654030 (Y.T.) and No.22540383 (M.S.). 


\begin{thebibliography}{99}
\bibitem{dxy}
When the symmetry of the singlet component of pair potential 
is $d_{xy}$-wave ($d_{x^2-y^2}$-wave),
the number of the sign change of the real or imaginary part of 
triplet one on the Fermi surface is two (six).
Thus, we call the mixed pair potential $d_{xy}+p$-wave ($d_{x^{2}-y^{2}} + f$-wave).
\end{thebibliography}
\end{document}